\documentclass[12pt]{article}
\usepackage{amssymb}

\usepackage{graphicx}
\usepackage{amsmath}

\begin{document}

\title{Casimir energy density in closed hyperbolic universes\thanks{%
Contribution to the \textit{Fifth Alexander Friedmann Seminar on Gravitation
and Cosmology}, Jo\~{a}o Pessoa, Brazil, April 2002.}}
\author{\textbf{Daniel M\"{u}ller} \\
\textit{Instituto de F\'{i}sica, Universidade de Bras\'{i}lia}\\
\textit{\ Bras\'{i}lia, DF, Brazil}\\
\textit{E-mail:} \texttt{muller@fis.unb.br}\\
and \and \textbf{Helio V. Fagundes} \\
\textit{Instituto de F\'{i}sica Te\'{o}rica, Universidade Estadual Paulista}%
\\
\textit{\ S\~{a}o Paulo, SP, Brazil}\\
\textit{\ E-mail:} \texttt{helio@ift.unesp.br}}
\maketitle

\begin{abstract}
The original Casimir effect results from the difference in the vacuum
energies of the electromagnetic field, between that in a region of space
with boundary conditions and that in the same region without boundary
conditions. In this paper we develop the theory of a similar situation,
involving a scalar field in spacetimes with negative spatial curvature.
\end{abstract}

\section{\protect\medskip INTRODUCTION}

In a previous work \cite{MFO} the Casimir energy density was obtained for a
Robertson-Walker (RW) cosmological model with constant, negative spatial
curvature. Its spatial section was Weeks manifold, which is the hyperbolic
3-manifold with the smallest volume (normalized to $K=-1$ curvature) in the 
\textsc{SnapPea} census \cite{snappea}.

Here we further develop and clarify the theoretical formalism of that paper.

Our sign conventions for general relativity are those of Birrell and Davies 
\cite{BiDa}: metric signature $(+---)$, Riemann tensor \ $R_{\ \beta \gamma
\delta }^{\alpha }=\partial _{\delta }\Gamma _{\ \beta \gamma }^{\alpha
}-... $ , Ricci tensor \ $R_{\mu \nu }=R_{\ \mu \alpha \nu }^{\alpha }.$

\section{THE ORIGINAL CASIMIR EFFECT}

The original effect was calculated by Casimir \cite{casimir}. Briefly, one
sets two metallic, uncharged parallel plates, separated by a small distance $%
a$. Between\ them the electromagnetic field wavenumbers normal to the plates
are constrained by the boundaries. So there is a difference $\delta E$
between the vacuum energy for this configuration and the vacuum energy for
unbounded space. If $A$ is the area of each plate, one has (see, for
example, \cite{IZ}, \cite{milonni}, \cite{BMM}) 
\begin{equation*}
\frac{\delta E}{A}\doteqdot \frac{\hslash c}{2}\int \int \frac{dk_{x}dk_{y}}{%
(2\pi )^{2}}\left[ \sum_{n\in Z}\sqrt{k_{x}^{2}+k_{y}^{2}+(\pi n/a)^{2}}\
-2a\int \frac{dk_{z}}{2\pi }\sqrt{k_{x}^{2}+k_{y}^{2}+k_{z}^{2}}\right] ,
\end{equation*}
where we omitted damping factors needed to avoid infinities. The results is $%
\ $ 
\begin{equation*}
\delta E(a)=-\frac{\pi ^{2}\hbar c}{720a^{3}}A\ 
\end{equation*}
for the energy difference, and 
\begin{equation*}
F(a)=-\frac{\pi ^{2}}{240a^{4}}A\ 
\end{equation*}
for the attractive force between the plates.

\section{CASIMIR ENERGY (CE) IN COSMOLOGY WITH NONTRIVIAL TOPOLOGY}

There is no boundary for a universe model with closed (i.e., compact and
boundless) spatial sections. But a field in these models has periodicities,
which leads to an effect similar to the above one, that may also be called a
Casimir effect.

A simple example, taken from Birrell and Davis \cite{BiDa}, is that of a
scalar field $\phi (t,x)$\ in spacetime $\mathbf{R}^{1}\times S^{1},$ with
one closed space direction. If $S^{1}$ has length $L$ then

\begin{equation*}
\phi (t,\ x+L)=\phi (t,x)\ ,
\end{equation*}
and the vacuum energy density is

\begin{equation*}
\rho =-\pi \hbar c/6L^{2}.
\end{equation*}

An analytical expression for the CE in a class of closed hyperbolic
universes (CHUs) was obtained by Goncharov and Bytsenko \cite{GB}.

Here we develop a formalism succintly described in \cite{MFO}, for the
numerical calculation of the CE \textit{density }of closed hyperbolic
universes.

\ 

Our notation: $i,j,...=1-3;\ \alpha ,\mu =0-3;\ \mathbf{x}=(x^{i});\
x=(x^{\mu })=(t,$ $\mathbf{x).}$

Sign conventions are those of \cite{BiDa}: metric signature $(+---)$,
Riemann tensor \ $R_{\ \beta \gamma \delta }^{\alpha }=\partial _{\delta
}\Gamma _{\ \beta \gamma }^{\alpha }-...$ , Ricci tensor: \ $R_{\mu \nu
}=R_{\ \mu \alpha \nu }^{\alpha }.$ \qquad \qquad \qquad \qquad \qquad \qquad

\qquad \qquad \qquad \qquad \qquad \qquad

\section{SCALAR FIELD $\protect\phi (x)$ IN CURVED SPACETIME}

The action for a scalar field in a curved spacetime of metric $g_{\mu \nu }$
and mass $m$ is \ 
\begin{equation*}
S=\int \mathcal{L}(x)d^{4}x\ ,
\end{equation*}
with 
\begin{equation*}
\mathcal{L=}\frac{1}{2}\sqrt{-g}\left[ g^{\mu \nu }\phi _{;\mu }\phi _{;\nu
}-(m^{2}+\xi R)\phi ^{2}\right] \ ,
\end{equation*}
where $R$ is scalar curvature of spacetime, $g=\det (g_{\mu \nu }),$ and $%
\xi $ is a constant.

With $\xi =1/6$ (``conformal'' value) we get the equation for $\phi (x)$: 
\begin{equation*}
\frac{\delta S}{\delta \phi }=0\Rightarrow (\square +m^{2}+\frac{1}{6}R)\phi
=0,
\end{equation*}
where $\square $ is the generalized d'Alembertian: 
\begin{equation*}
\square \phi =g^{\mu \nu }\triangledown _{\mu }\triangledown _{\nu }\phi
=(-g)^{-1/2}\partial _{\mu }\left[ (-g)^{1/2}g^{\mu \nu }\partial _{\nu
}\phi \right] .
\end{equation*}

The energy-momentum tensor is (cf. \cite{BiDa}) \ 
\begin{eqnarray*}
T_{\mu \nu } &=&2(-g)^{-1/2}\ \delta S/\delta g^{\mu \nu } \\
&=&\frac{2}{3}\phi _{;\mu }\phi _{;\nu }+\frac{1}{6}g_{\mu \nu }\phi
_{;\sigma }\phi ^{;\sigma }-\frac{1}{3}\phi _{;\mu \nu }+\frac{1}{12}g_{\mu
\nu }\phi \square \phi \\
&&-\frac{1}{6}R_{\mu \nu }\phi ^{2}+\frac{1}{24}g_{\mu \nu }R\phi ^{2}+\frac{%
1}{4}g_{\mu \nu }m^{2}\phi ^{2}\ .
\end{eqnarray*}

\medskip

\section{COORDINATES IN \textit{H}$^{3}$}

The hyperbolic (or B\'{o}lyai-Lobachevsky) space $H^{3}$ is isometric to the
hypersurface 
\begin{equation*}
(x^{4})^{2}-\mathbf{x}^{2}=1,\ x^{4}\geq 1\ ,
\end{equation*}
imbedded in an abstract Minkowski space $(\mathbf{R}^{4},diag(1,1,1,-1)).$\
\ 

This upper branch of a hyperboloid is similar to\ the mass shell of particle
physics, \ 
\begin{equation*}
E^{2}-\mathbf{p}^{2}=m^{2},\ E\geq m\ .
\end{equation*}
\ 

Hence a point in $H^{3}$ may be represented by the Minkowski coordinates $%
x^{b},\ b=1-4,$ subject to constraints (1), and rigid motions in $H^{3}$ are
proper, orthochronous Lorentz transformations.

We relate the spherical coordinates $(\chi ,\theta ,\varphi )$ to the
displaced Minkowski ones $x^{b}-x^{\prime b},\ b=1-4:$

\begin{center}
\begin{tabular}{l}
$x^{1}-x^{\prime 1}=\sinh \chi \sin \theta \cos \varphi \ ,$ \\ 
$x^{2}-x^{\prime 2}=\sinh \chi \sin \theta \sin \varphi \ ,$ \\ 
$x^{3}-x^{\prime 3}=\sinh \chi \cos \theta \ ,$ \\ 
$x^{4}-x^{\prime 4}=\cosh \chi \ .$%
\end{tabular}
\end{center}

Note that $\chi (\mathbf{x,x}^{\prime })=\sinh ^{-1}|\mathbf{x-x}^{\prime
}|\ .$

\section{STATIC MODELS OF NEGATIVE SPATIAL CURVATURE}

The Robertson-Walker metric for spatial curvature $K=-1/a^{2}$ is\ 
\begin{eqnarray*}
ds^{2} &=&dt^{2}-a^{2}(d\chi ^{2}+\sinh ^{2}\chi \ d\Omega ^{2}) \\
&=&dt^{2}-a^{2}\left( \delta _{ij}-\frac{x^{i}x^{j}}{1+\mathbf{x}^{2}}%
\right) dx^{i}dx^{j}\ ,
\end{eqnarray*}
where in general $a=a(t).$

Einstein's equations give\ 
\begin{eqnarray*}
\left( \frac{\dot{a}}{a}\right) ^{2} &=&\frac{1}{a^{2}}+\frac{8\pi G}{3}\rho
+\frac{\Lambda }{3}\ , \\
\frac{3\ddot{a}}{a} &=&-4\pi G(\rho +3P)+\Lambda \ .\ 
\end{eqnarray*}

Assuming $\dot{a}$ $=$ $\ddot{a}=0$ and $P=\rho /3$ we get $%
a^{2}=-3/2\Lambda ,$ hence $\Lambda <0,$ and\ \ \ 
\begin{equation*}
a=\sqrt{3/2|\Lambda |\ },
\end{equation*}
\begin{equation*}
\rho =\Lambda /8\pi G<0\ .
\end{equation*}

We will comment below on this negative energy density.

These models are stable (!) under curvature fluctuations: 
\begin{equation*}
a\rightarrow a+\varepsilon (t)\Longrightarrow \ddot{\varepsilon}+|\Lambda
|\varepsilon =0\ .
\end{equation*}

\section{\ CLOSED HYPERBOLIC 3-MANIFOLDS (CHMs)}

A CHM is obtained by a pairwise identification of the $n$ faces of a \textit{%
fundamental polyhedron (FP)}, or \textit{Dirichlet domain}, in hyperbolic
space. It is isometric to the quotient space $H^{3}/\Gamma ,$ where $\Gamma $
is a discrete group of isometries of $H^{3},$ defined by generators\textit{\ 
}and relations, which acts on $H^{3}$ so as to produce the \textit{%
tesselation} 
\begin{equation*}
H^{3}=\underset{\gamma \in \Gamma }{\cup }\gamma (FP)\ .
\end{equation*}

Each cell $\gamma (FP)$ is a copy of $FP$, hence we have periodicity of
functions on a CHM, and the possibility of a cosmological Casimir effect.

\textit{Face-pairing }generators $\gamma _{k},\ k=1-n,$ satisfy 
\begin{equation*}
FP\cap \gamma _{k}(FP)=\text{ face }k\text{ of }FP\text{ .}
\end{equation*}
With these generators the relations also have a clear geometrical meaning:
they correspond to the \textit{cycles }of cells around the edges of $FP.$

The software \textsc{SnapPea }\cite{snappea} includes a ``census'' of about
11,000 orientable CHMs, with normalized volumes from 0.94270736 to
6.45352885. For each of these the $FP$ centered on a special basepoint $O$
is given, as well as the face-pairing generators in both the $SL(2,C)$ and
the $SO(1,3)$ representations.

An algorithm \cite{jwpc} to find a set of \ cells $\gamma (FP)$ that
completely cover a ball of radius $r$ reduces this problem to one of finding
all motions $\gamma \in \Gamma ,$\ such that 
\begin{equation*}
\text{distance}[O,\gamma (O)]<r+(\text{radius of }FP\text{'s circumscribing
sphere})\ .
\end{equation*}

For a study of CHMs from a cosmological viewpoint, see for example \cite
{LaLu} and references therein. For numerical data on a couple of them, see 
\cite{QGAII}, \cite{SUN}.

\section{CLOSED HYPERBOLIC UNIVERSES}

We are considering \textit{static} CHUs. As obtained in Sec. 6, the metric
is 
\begin{equation*}
ds^{2}=dt^{2}-\frac{3}{2|\Lambda |}\left( \delta _{ij}-\frac{x^{i}x^{j}}{1+%
\mathbf{x}^{2}}\right) dx^{i}dx^{j}\ .
\end{equation*}

The spacetimes have nontrivial topology:\ \ 
\begin{equation*}
M^{4}=\mathbf{R}^{1}\mathbf{\times \Sigma \ ,}
\end{equation*}
where $\mathbf{R}^{1}$ is the time axis and $\Sigma =H^{3}/\Gamma $\ is a
CHM.\ \ 

As found above, these models have negative energy density, $\rho =\Lambda
/8\pi G,$ which has no obvious physical meaning, and violates the energy
condition $T_{\mu \nu }u^{\mu }u^{\nu }\geq 0.$  But we are dealing with the
very early universe, where one feels freer to speculate. And a recent paper
by Olum \cite{olum} casts doubt on the universality of this condition. 

Our original motivation was the possibility of preinflationary
homogenization through chaotic mixing, leading to $\Omega _{0}<1$ inflation
(cf. Cornish et al. \cite{cornish}). \ 

Another guess is that these models might have a place in the path integrals
for quantum cosmology.

\label{textMid}

\section{ THE ENERGY-MOMENTUM OPERATOR}

If we use use the equation for $T_{\mu \nu }$ in Sec. 4 to calculate $%
<0|T_{\mu \nu }|0>$ we get terms like 
\begin{equation*}
<0|\phi (x)\phi (x)|0>\ ,
\end{equation*}
which lead to infinities.

To avoid this one replaces $x$ by $x^{\prime }$ in the first factor, then in
the second factor, and average the result. Thus the above expectation value
becomes one-half Hadamard's function $G^{(1)}:$ \ 
\begin{equation*}
G^{(1)}(x,x^{\prime })=\ <0|[\phi (x),\ \phi (x^{\prime })]_{+}|0>\ ,
\end{equation*}
and we obtain (cf. Christensen \cite{christensen}, with our signs) 
\begin{equation*}
<0|T_{\mu \nu }(x,x^{\prime })|0>\ =\hat{T}_{\mu \nu }(x,x^{\prime
})G^{(1)}(x,x^{\prime })\ ,\ 
\end{equation*}
with the operator\ 

\begin{eqnarray*}
&&\hat{T}_{\mu \nu }(x,x^{\prime })=\frac{1}{6}(\nabla _{\mu }\nabla _{\nu
^{\prime }}+\nabla _{\mu ^{\prime }}\nabla _{\nu })+\frac{1}{12}g_{\mu \nu
}(x)\nabla _{\rho }^{\ \ \ }\nabla ^{\rho ^{\prime }} \\
&&-\frac{1}{12}(\nabla _{\mu }\nabla _{\nu }+\nabla _{\mu ^{\prime }}\nabla
_{\nu ^{\prime }})+\frac{1}{48}g_{\mu \nu }(x)(\nabla _{\rho }^{\ \ \
}\nabla ^{\rho }+\nabla _{\rho ^{\prime }}^{\ \ \ }\nabla ^{\rho ^{\prime }})
\\
&&-\frac{1}{12}\left[ R_{\mu \nu }(x)-\frac{1}{4}g_{\mu \nu }(x)R(x)\right] +%
\frac{1}{8}m^{2}g_{\mu \nu }(x)\ ,
\end{eqnarray*}
where $\nabla _{\alpha }$ and $\nabla _{\alpha ^{\prime }}$ are covariant
derivatives with respect to $x^{\alpha }$ and $x^{\prime \alpha }$,
respectively.

Eventually one takes the limit $x\rightarrow x^{\prime }$ to get the CE\
density. But first we have to investigate $G^{(1)}(x,x^{\prime })$.

\ 

\section{\protect\medskip FEYNMAN'S PROPAGATOR}

\ $G^{(1)}(x,x^{\prime })$ will be obtained from Feynman's propagator for a
scalar field $G_{F}(x,x^{\prime }).$

In an $\mathbf{R}^{1}\times H^{3}$ universe$,$ $G_{F}$ gets an extra factor $%
(\chi /\sinh \chi ),$ where $\chi =\sinh ^{-1}|\mathbf{x-x}^{\prime }|$,
with respect to its flat spacetime counterpart; and the squared interval $%
(x-x^{\prime })^{2}$ in the latter becomes $2\sigma =(t-t^{\prime
})^{2}-a^{2}\chi ^{2},$ which is the squared geodesic distance between $x$
and $x^{\prime }$. \qquad The derivation of the following expression (with
opposite sign because of a different metric signature) is outlined in \cite
{MFO}: \ 
\begin{equation*}
G_{F}(x,x^{\prime })=\frac{m^{2}}{8\pi }\frac{\chi }{\sinh \chi }\frac{%
H_{1}^{(2)}\left( m\sqrt{2\sigma }\right) }{m\sqrt{2\sigma }}\ ,
\end{equation*}
where $H_{1}^{(2)}$ is Hankel's function of second kind and degree one.

For our spacetime $\mathbf{R}^{1}\times H^{3}/\Gamma ,$ point $\mathbf{x}$\
may be reached by the projections of all geodesics that link $\mathbf{x}%
^{\prime }$ to $\gamma \mathbf{x}$ in the covering space $H^{3}$ [in
Minkowski coordinates $(\gamma \mathbf{x})^{i}=\sum_{b=1}^{4}\gamma _{\
b}^{i}x^{b},\ i=1-3$].

Therefore our propagator is \ \ 
\begin{equation*}
G_{F}(x,x^{\prime })=\frac{m^{2}}{8\pi }\sum_{\gamma \in \Gamma }\frac{\chi
(\gamma )}{\sinh \chi (\gamma )}\dfrac{H_{1}^{(2)}(m\sqrt{2\sigma (\gamma )})%
}{m\sqrt{2\sigma (\gamma )}}\ ,
\end{equation*}
with $\chi (\gamma )=\sinh ^{-1}|\gamma \mathbf{x-x}^{\prime }|$ and $%
2\sigma (\gamma )=(t-t^{\prime })^{2}-a^{2}\chi ^{2}(\gamma )$. \ 

\section{\protect\medskip HADAMARD'S FUNCTION}

\medskip Hadamard's function is related to $G_{F}$ and the principal value
Green's function $\bar{G}$ by 
\begin{equation*}
G_{F}(x,x^{\prime })=-\bar{G}(x,x^{\prime })-\frac{i}{2}G^{(1)}(x,x^{\prime
})\ .
\end{equation*}
In our problem, both $\bar{G}$ and $G^{(1)}$ are real, so that 
\begin{equation*}
G^{(1)}(x,x^{\prime })=-2\text{ Im }G_{F}(x,x^{\prime })\ .
\end{equation*}

We need $G^{(1)}(x,x^{\prime })$ for $x^{\prime }$ near $x,$ hence when $%
2\sigma (\gamma )$ is near $-a^{2}\chi ^{2}\leq 0$. So we write the argument
of $H_{1}^{(2)}$ as $iu_{\gamma },$ with $u_{\gamma }=m\sqrt{2|\sigma
(\gamma )|}.$ From the properties of Bessel functions, 
\begin{equation*}
-2\ \text{Im }\left[ (iu_{\gamma })^{-1}H_{^{1}}^{(2)}(iu_{\gamma })\right]
=(4/\pi )u_{\gamma }^{-1}K_{1}(u_{\gamma })\ ,
\end{equation*}
where $K_{1}$ is a modified Bessel function of degree one. \ 

Hadamard's function for a universe $\mathbf{R}^{1}\times H^{3}/\Gamma $ is
then 
\begin{equation*}
G_{\Gamma }^{(1)}{\LARGE (}x,x^{\prime }{\LARGE )}=\frac{m^{2}}{2\pi ^{2}}%
\sum_{\gamma \in \Gamma }\frac{\chi (\gamma )}{\sinh \chi (\gamma )}\frac{%
K_{1}\left( u_{\gamma }\right) }{u_{\gamma }}\ ,
\end{equation*}

The $\gamma =1$ term in this sum corresponds to the infinite $\mathbf{R}%
^{1}\times H^{3}$ universe. Similarly to what was done for the two-plate
Casimir effect in Sec. 2, we subtract it out to get a finite energy density.
Therefore the expression in Sec. 9 for $<0|T_{\mu \nu }(x,x^{\prime })|0>$
leads to 
\begin{equation*}
<0|T_{\mu \nu }(x,x^{\prime })|0>_{C}\ =\hat{T}(x,x^{\prime
})G_{C}(x,x^{\prime })\ ,
\end{equation*}
where $G_{C}=G_{\Gamma }^{(1)}-G_{\{1\}}^{(1)}.$ \ 

\ \ 

\section{THE CASIMIR ENERGY DENSITY}

Finally, the CE density is given by 
\begin{equation*}
<0|T_{00}(\mathbf{x})|0>_{C}\ =\lim_{x^{\prime }\rightarrow x}\hat{T}%
_{00}(x,x^{\prime })\ G_{C}(x,x^{\prime })\ ,
\end{equation*}
where

\begin{equation*}
G_{C}{\LARGE (}x,x^{\prime }{\LARGE )}=\frac{m^{2}}{2\pi ^{2}}\sum_{\gamma
\in \Gamma -\{1\}}\frac{\sinh ^{-1}|\gamma \mathbf{x-x}^{\prime }|}{|\gamma 
\mathbf{x-x}^{\prime }|}\frac{K_{1}(m\sqrt{-2\sigma (\gamma )})}{m\sqrt{%
-2\sigma (\gamma )}}\ ,
\end{equation*}
with $-2\sigma (\gamma )=a^{2}(\sinh ^{-1}|\gamma \mathbf{x-x}^{\prime
}|)^{2}-(t-t^{\prime })^{2}.$ \ 

Looking at the expressions for $\hat{T}_{00}(x,x^{\prime })$ and $%
G_{C}(x,x^{\prime }),\ $one sees they are pretty complicated.\ \ 

Now enters the power of computers!\ \ \ 

Calculations were performed by one of us (DM), for a grid of points $(\theta
,\varphi )$ on a sphere of radius $r$ inside the $FP$, for a number of
static CHUs. In \cite{MFO} the parameters are, in Planckian units, $m=0.5,\
a=10,$ and $r=0.6$, and the summation for $G_{C}(x,x^{\prime })$ contains a
few thousand terms; the obtained density values oscillate around $%
-2.65\times 10^{-6}.$ New results will be published elsewhere. \ \ 

\medskip\ 

We thank FAPESP and CNPq for partial financial help. HVF thanks George
Matsas for conversations on field theory in curved spacetime.

\label{textEnd}


\medskip

\begin{thebibliography}{99}
\bibitem{MFO}  D. M\"{u}ller, H. V. Fagundes, and R. Opher, Physical Review
D63, 123508 (2001)

\bibitem{snappea}  J. R. Weeks, \textit{Snappea: a computer program for
creating and studying three-manifolds}, freely available at Web site \texttt{%
www.northnet.org/weeks}

\bibitem{BiDa}  N. D. Birrell and P. C. W. Davies, \textit{Quantum Fields in
Curved Space} (Cambridge University Press, Cambridge, 1982)

\bibitem{casimir}  H. B. G. Casimir, Proc. Kon. Ned. Akad. Wet. 51, 793
(1948)

\bibitem{IZ}  C. Itzykson and J.-B. Zuber, \textit{Quantum field theory (}%
Addison-Wesley, New York, 1980)

\bibitem{milonni}  P. W. Milonni, \textit{The quantum vacuum: an
introduction to quantum electrodynamics} (Academic, Boston, 1994) \ 

\bibitem{BMM}  M. Bordag, U. Mohideen, and V. M. Mostepanenko, Phys. Rep.
353, 1 (2001), Sec. 2.2

\bibitem{GB}  Yu. P. Goncharov and A. A. Bytsenko, Class. Quant. Grav. 8,
1211 (1991)

\bibitem{LaLu}  M. Lachi\`{e}ze-Rey and J.-P. Luminet, Phys. Rep. 254, 135
(1995)

\bibitem{QGAII}  H. V. Fagundes, Astrophys. J. 338, 618 (1989); Errata in
Astrophys. J. 349, 678 (1990)

\bibitem{SUN}  H. V. Fagundes, Phys. Rev. Lett. 70, 1579 (1993)

\bibitem{jwpc}  J. R. Weeks, private communication

\bibitem{olum}  K. D. Olum, gr-qc/0205134

\bibitem{cornish}  N. J. Cornish, D. N. Spergel, and G. D. Starkman, Phys.
Rev. Lett. 77, 215 (1996)

\bibitem{christensen}  S. M. Christensen, Phys. Rev. D17, 946 (1978) \ \ 
\end{thebibliography}
\end{document}